# Simulation of Heterogeneous Atom Probe Tip Shapes Evolution during Field Evaporation Using a Level Set Method and Different Evaporation Models

Zhijie Xu [a)], Dongsheng Li, Wei Xu, Arun Devaraj [a)], Robert Colby, Suntharampillai Thevuthasan

Pacific Northwest National Laboratory, Richland, WA 99352

B. P. Geiser and D. J. Larson

CAMECA Instruments Inc., Madison, WI 53711

**Abstract:**

In atom probe tomography (APT), accurate reconstruction of the spatial positions of field evaporated ions from measured detector patterns depends upon a correct understanding of the dynamic tip shape evolution and evaporation laws of component atoms. Artifacts in APT reconstructions of heterogeneous materials can be attributed to the assumption of homogeneous evaporation of all the elements in the material in addition to the assumption of a steady state hemispherical dynamic tip shape evolution. A level set method based specimen shape evolution model is developed in this study to simulate the evaporation of synthetic layered-structured APT tips. The simulation results of the shape evolution by the level set model qualitatively agree with the finite element method and the literature data using the finite difference method. The asymmetric evolving shape predicted by the level set model demonstrates the complex evaporation behavior of heterogeneous tip and the interface curvature can potentially lead to the artifacts in the APT reconstruction of such materials. Compared with other APT simulation methods, the new method provides smoother interface representation with the aid of the intrinsic

---

[a)] Corresponding author contact information: zhijie.xu@pnnl.gov
Corresponding author contact information: arun.devaraj@pnnl.gov

sub-grid accuracy. Two evaporation models (linear and exponential evaporation laws) are implemented in the level set simulations and the effect of evaporation laws on the tip shape evolution is also presented.

**Introduction**

Atom probe tomography (APT) is widely used for nanoscale 3D compositional characterization of a variety of materials ranging from metals and alloys [1-7], dielectrics[8, 9], semiconductors[10, 11] to biological[12, 13] and geological materials[14]. The past decade has seen a dramatic development of APT instrumentation/data collection and sample preparation methods [15]. On the data collection front, local electrode and laser pulsing have been introduced widening the application of APT beyond conductive materials. On the sample preparation front, focused ion beam (FIB) based site specific sample preparation methods have been introduced [16-18] permitting a better control of region selection for APT analysis as well as facilitating APT specimen preparation of nonconductive materials. However a new challenge surfaces when the material systems analyzed by APT expand from homogeneous metallic system to heterogeneous materials: the traditional APT data reconstruction protocol introduces artifacts due to the underlying assumption of uniform and isotropic evaporation.

The APT data reconstruction process is influenced by two distinct physical events that occur during APT analysis: field evaporation of the ions from the surface of an APT sample and projection of these evaporated ions to a position sensitive detector. This work will focus on the first step (evaporation process) which is a complex function of atom properties, atomic environment and emitter geometry. Traditional APT reconstruction algorithm for generating the 3D atomic position from the detector positions of ions is limited by the assumptions of homogeneous structure, uniform evaporation rate for different components and specific tip geometry [19]. The currently used reconstruction algorithm is modified from the method proposed by Bas et al. [19], and it is based on the assumption that the tip maintains a constant hemispherical shape during evaporation [20]. The apparent field strength $F$ at the apex is determined by an approximation solution [21]:

$$F = \frac{U}{k \cdot r} \tag{0}$$

where U is the applied voltage, r is the radius of curvature and the empirical parameter k is the geometric factor/field factor accounting for the influence of the shaft and the geometry of the surrounding vacuum chamber [22]. During APT analysis to keep the ion detection rate constant, the applied voltage will be adjusted according to the increase in tip radius with evaporation. Using a simplified geometry model with the assumption of uniform hemispherical tip shape evolution, analytical solutions of the geometric factor has been derived in several models. These reconstruction protocols with analytical solutions of emitter evaporation and point projection have been applied in currently available data analysis tools. Generally, the analysis results agree well with simulation results and experimental data validated by other instrumentation. However, this is not always the case in a heterogeneous system consisting of phases with different evaporation behavior. For most complex materials systems, the traditional reconstruction method will create global distortions in the final reconstructed data due to the magnification artifacts, which is mainly caused by the evaporation induced non-uniform apex shapes. Understanding such artifacts and improving the data analysis to account for such effects is very important to the further successful application of APT.

One exemplary illustration of the local magnification due to the heterogeneity is in the case of MgO embedded with gold particles [23, 24]. Before evaporation, the APT needle sample has a smooth surface with gold particles embedded in the MgO matrix. After the evaporation, gold nanoparticles protrude out at the surface due to the slow evaporation rate, in contrast to the MgO matrix. In these experiments, the 3-D reconstruction analysis was carried out with the assumptions described above and the complex nature of the surface due to the Au particle protrusion from the surface contributes to trajectory aberrations. There has also been past experimental evidence provided by performing TEM imaging after interrupted APT analysis of multilayer samples highlighting the non-hemispherical tip geometry changes during the evaporation of layers with differing evaporation fields[25]. Recent APT-TEM correlation results on vertical Si-SiO$_2$ interfaces also highlight the non-hemispherical tip shape evolution during APT analysis of heterogeneous materials[26]. In addition to the experimental evidence, a number of simulation efforts has been focused on the evaporation of hemispherical APT tips consisting of layers or precipitates with varying evaporation fields [27-31].

To take full advantage of the high resolution inherited in atom probe tomography, a new efficient and accurate tip evaporation model is crucial for next generation atomic structure reconstruction software [25, 32-35]. In most tip evaporation models, finite difference or finite element models with atomic lattice structure is used to simulate the geometric tip evaporation and the following trajectory projection. Larson et al. [36] used a finite difference method to simulate the dynamic tip shape evolution during evaporation of a multilayer-structured tip, where each layer material had different evaporation field strength. Three different initial APT tip geometries were considered including two conditions where the interface between a high evaporation field layer material and low evaporation field material layer was considered perpendicular to the axis of the tip and one case where the interface between the same layers were considered to be parallel to the axis of the APT sample. In all three simulation cases, the interface structures in the reconstructed data were distorted due to non-hemispherical tip shape evolution during evaporation. This necessitated additional corrections to be imposed on the standard reconstruction protocol in order to obtain an accurate interface definition representative of the initial stage of the APT tip. There have also been finite element simulations aimed at understanding evaporation of heterogeneous composite materials consisting of nanoscale precipitates with widely different evaporation field compared to the matrix [37-39]. Pseudo 3D simulation models with revolution axisymmetric geometry have also been developed to reduce the computational expense [40-42]. Due to the large size of the sample compared with the atomic lattice and the size of APT chamber compared with the tip, the simulation can be extremely computationally expensive. To make it affordable in most models, the tip size is shrunk and the distance between the electrode and tip is shortened, where computationally efficient models are needed to overcome these difficulties.

To improve the computational efficiency, numerical simulations require more efficient approaches. In the last several decades, level set, phase field and many other interface tracking methods have been widely applied to solve the dynamic moving interfaces. Examples are the application of the phase field and level set methods to simulate solute precipitation/dissolution problem with evolving solid–liquid interfaces [43-46] and solid-solid interfaces [47-49]. Both methods have subgrid scale accuracy through linear interpolation such that the interface can be more accurately represented while interface normals and curvature can be also conveniently calculated. The phase-field approach does not require explicit computing of the moving interface so that numerical difficulties associated with the moving interface are avoided and no explicit interface tracking is required. However, phase field approach requires a complex asymptotic analysis in order to find the relation between parameters of the sharp-interface model and the phase field model [50, 51]. A level set approach does not require such an asymptotic analysis and can naturally handle the discontinuities at the interface through explicit interface tracking. Haley et al. [52] have applied level set method for modeling homogeneous tip evaporation where curvature drive flow was used to mimic the tip evolution. In this study, the level set method is used to simulate the field evaporation of heterogeneous specimen coupling with a simultaneously updated full Poisson solution for electric potential field.

**METHODS DESCRIPTION**

To validate the level set model developed for APT specimen evaporation, we compared the results with simulation results from the finite element and the finite difference method. Both methods solve the Poisson equation to retrieve potential field and local electric field, and then implement an evaporation process. The process is then iterated until stopped at a preset critical value of either tip height or evaporated ion volume/number.

**Description of Level Set Method**

In our approach, a level set based evaporation model is developed to simulate the evolution of tip surface morphology during the evaporation of multi-layered heterogeneous materials. Current evaporation models for APT utilize finite difference or finite element methods where the staircase representation is commonly used to describe the tip surface. The accuracy of tip geometry cannot be higher than the resolution of grid cell. Such methods are also not straightforward when the interface curvature and normals are involved in the calculation for atomic evaporation.

The level set method was originally introduced for solving multiphase flow problems involving dynamic and complex interface evolution between two phases. Historically these "moving boundary and/or interface" problems have been very challenging from a computational point of view. Level set methods [53] are based on the tracking or capturing of sharp interfaces, while complex geometries and morphology changes can be handled efficiently. They have been applied to interface tracking in the areas of computational fluid dynamics and fracture propagation in the framework of extended finite element method. For simulation of field evaporation, we use level set method to investigate the dynamic evolution of the interface between two distinct phases, namely the vacuum phase and the material phases. A continuous level set function $\phi(x)$ is introduced in this model. The zero-level of the level set function denotes the exact position of the interface between vacuum and material phases, i.e.

$$\Gamma = \{\mathbf{x} | \phi(\mathbf{x},t) = 0\} \tag{0}$$

where level set function $\phi$ is defined as a signed distance from the interface $\Gamma$, (positive in vacuum phase and negative in the material phase. The interface $\Gamma$ is implicitly represented by the zero-level of level set field. The evolution of entire level set field is numerically computed based on hyperbolic conservation laws,

$$\frac{\partial \phi}{\partial t} + \mathbf{V} \cdot \nabla \phi = 0 \tag{0}$$

where $\mathbf{V}$ is the interface moving velocity defined on the interface $\Gamma$. The local interface normal $\mathbf{n}$ and curvature can be conveniently calculated as

$$\mathbf{n} = \nabla \phi / |\nabla \phi| \text{ and } \kappa = \nabla \cdot (\nabla \phi / |\nabla \phi|) \tag{0}$$

In order to solve Eq. (3) numerically, the interface moving velocity $\mathbf{V}$ should be explicitly defined based on physical laws. Specifically for APT modelling, we first solve the Poisson equation

$$\nabla \cdot (\varepsilon_r(\mathbf{x}) \nabla \varphi) = 0 \tag{0}$$

for the electrical potential field $\varphi(\mathbf{x})$, where $\varepsilon_r(\mathbf{x})$ is a material dependent relative dielectric permittivity. Recently Vurpillot et. al. and Arnoldi et. al. have shown that a strict adherence to simple dielectric models fails to explain some important features of experimental atom probe data from bulk oxides or dielectric materials, and suggested residual conductivity due to internal or surface defects or photo carrier generation as the cause for this deviation[42, 54]. Inclusion of this residual conductivity into our model is deferred for a later publication. However we believe there is value in this relative dielectric permittivity based model, since this can serve as an initial model for materials with less resistivity. Based on the evaporation physics [22], the interface moving velocity along the normal direction can be defined as

$$V = a v_0 \exp\left(-Q(F)/k_B T\right) \tag{0}$$

where $a$ is the lattice constant, $v_0$ is the vibration frequency, $k_B$ is the Boltzmann constant and $T$ is the temperature. $Q(\mathbf{F})$ is the energy barrier with external electric field $\mathbf{F} = \nabla \varphi$ that is the gradient of the potential field. A linear approximation can be used to describe the dependence of $Q$ on the field vector where

$$Q(\mathbf{F}) = Q_0 \left(1 - F_1/F_0\right) \tag{0}$$

where

$$F_1 = \left(\left.\frac{\partial \varphi}{\partial n}\right|^+ + \left.\frac{\partial \varphi}{\partial n}\right|^-\right)\Big/ 2 \tag{0}$$

is the average local electric field along the normal direction, where $\left(\partial \varphi/\partial n\right)\big|^+$ is the field $\mathbf{F}$ on the vacuum phase side at the interface and $\left(\partial \varphi/\partial n\right)\big|^-$ is the value of field $\mathbf{F}$ along the normal direction on the material side. $Q_0$ is the energy barrier without any external field and $F_0$ is the material evaporation field strength.. Substitution of Eq. (7) into Eq. (6) leads to the exponential evaporation law where,

$$V = a v_0 \exp\left(-\alpha\left(1 - F_1/F_0\right)\right) = V_0 \exp\left(-\alpha\left(1 - F_1/F_0\right)\right) \tag{9}$$

Here $\alpha$ is a dimensionless parameter defined as $\alpha = Q_0/(k_B T)$, $V_0 = a v_0$ is the evaporation rate when $F_1 = F_0$. With $\alpha \approx 1$ and $F_1 \approx F_0$, the exponential evaporation law is reduced to the linear evaporation law,

$$V = a v_0 \left(F_1/F_0\right) \tag{10}$$

Both exponential (Eq. (9)) and linear (Eq. (10)) evaporation laws are implemented in the current level set models. The effect of two different evaporation laws on tip shape evolution is also investigated.

It is worth mentioning a level set model proposed by Haley et al. for homogeneous tip evaporation [52]. By substitution of the approximate electrical field F (Eq. (1)) into the linear evaporation law (Eq. (10)), the interface moving velocity can be written as:

$$V = \frac{a v_0 U}{k F_0} \kappa \tag{11}$$

where $\kappa = 1/r$ is the apparent curvature of the interface. Therefore, the APT problem is reduced to the curvature flow problem where $V$ is proportional to the local curvature. The curvature flow problem can be readily solved using a level set algorithm. In the present study, the Poisson equation (Eq. (5)) is explicitly solved instead of using the approximate electrical field (Eq. (1)).

**Description of Finite Difference and Finite Element Methods**

The simulation of field evaporation through finite difference and finite element methods generally follows the pioneering work developed by Vurpillot et al[28], Oberdorfer et. al.[38], Haley et al. [32], Gault et al. [55], Larson et al. [36], and Geiser et al. [56], in which the tip is discretized into stacks of elemental cells with each cell representing a single atom. Conservation of electric flux that can be expressed in the differential form as Laplace's equation $\nabla \cdot (\epsilon \nabla \varphi) = 0$ is numerically solved within the atom lattices, where $\epsilon$ is the dielectric constant, and $\varphi$ is the electrical potential that is computed at each point. The electric field on the tip surface can be computed as $F = -\nabla \varphi$.

In order to take into account the complexities introduced by the embedded heterogeneities within alloys or composites, a material dependent parameter $F_0$ (field evaporation strength) has been adopted to evaluate the corresponding probability for field evaporation as $p(F) = |F|/F_0$. Therefore, the atom with the highest electric field tends to have the highest evaporation probability and is determined to be the first field-evaporated specie. The cell that represents its atom volume is then explicitly dealt with, for example, converted to a vacuum cell, to reflect the incremental tip shape change. Recalculating the electrical field distribution with new geometry and repeating above procedures will continue the evaporation. The gradual change of the tip morphology can then be simulated.

**RESULTS AND DISCUSSION**

Results of the level set evaporation model were qualitatively compared to the results by using other numerical methods. Two sets of comparison were performed for the purpose of demonstration. One is to use our in-house developed model to compare with the level set method for a simulated tip with embedded spherical precipitate. The other set is to compare the level set results with finite difference simulations by Larson et al. [36] on a heterogeneous tip containing an interface between regions of high and low evaporation field materials.

**Comparison to finite element method**

FEM simulation is based on the general scheme developed by Oberdorfer et al. [38]. The numerical scheme is divided into consecutive stand-alone tasks to exploit the most computational efficiency. The calculation of the electric potential field is performed using the commercial FE software ABAQUS, where two dimensional 4-node bilinear piezoelectric element CPE4E is applied, while the automated data manipulation to resemble the dynamic field evaporation process, as well as the post-processing visualization, is realized through the mathematical package MATLAB®. The present FEM calculation was used to simulate a planar 2D synthetic APT tip for purpose of demonstration.

Similar to Oberdorfer's geometry and dimension [38], the sample tip is modelled as a hemispherical cup with 25nm radius. A fixed voltage is applied to the tip bottom. A grounded counter electrode with hemispherical shape is assumed 25nm away on top of the tip, as shown in Figure 1 (a). Note that this is for demonstration and is too short to reflect the actual electrode position in real experiment. A precipitate embedded in the matrix is also shown in Figure 1(a). The evaporation field strength for precipitate (material B) is assumed to be twice of that of the matrix material A. Each FEM element represents about 0.25 nm$^2$ area.

Figures 1(b)-(e) compares the simulated tip shape at different stages using the FEM and level set methods. Figures 1(b) and (c) show tip shapes after 23% material evaporated and Figs. 1(d) and (e) show tip shapes after 25% material evaporated. These intermediate steps are chosen to highlight the effect of non-hemispherical tip shape evolution when the embedded precipitate protrudes out of the surface of the APT specimen. Both simulations were carried out on a commercial desktop computer with 2.90GHz CPU and 8GB RAM. Both methods capture the major feature of high field embedded particles protruding out during field evaporation and are qualitatively similar to previous simulated results of precipitate evaporation [37, 38, 57, 58]. On the other side, simulation results from the level set method demonstrate a smooth tip surface, compared with the simulation results from FEM. The surface from FEM simulation is obviously rougher than that from the level set method due to the staircase representation of the interface. To achieve the same resolution of representation for surface, FEM would have to have a much smaller mesh that leads to a higher computational expense. Figures 1(f) and (g) illustrate the electric potential field contours by FEM and the level set methods. The surface geometry and the electrical potential field contours by both methods are very similar to each other. Our preliminary comparison found that the level set model has a potential gain of computational efficiency as it takes 8 to 10 hours for a computational domain of 120x120=14,400 grids in contrast to ~20 hours for the FEM simulation (~15,000 4-node quad elements) to complete the same evaporation process. More rigorous studies are needed to quantitatively compare the computational efficiency for both methods.

**Comparison to finite difference method**

Similar to the specimen geometry used by Larson et. al. [29], three different initial APT specimen geometries were considered including two cases with interfaces oriented perpendicular to APT specimen axis with (a) a low field material layer A on a high field material layer B; (b) a high field material layer B on a low field material A; and a third case (c) a high field material B on the side of a low field material A with interface parallel to APT specimen axis. The geometry configuration and dimensions are shown in Figure 2(a), (b) and (c).

Similar to the finite difference method, the material B is considered to have approximately 20% higher field strength than that of material A. The entire tip in Figure 2(a) and (b) is a circle with a radius of 18.4 nm positioned on a 9.6 nm×36.8 nm rectangle. The geometry in Figure 2(c) has half material A on the left side and half material B on the right side with a similar overall tip geometry like Figures 2(a) and (b). The entire simulation domain of (96 nmx96 nm) for Figure 2 has a grid cell size of 0.8nm. In level set simulation, the material *A* has a lower non-dimensional field strength of 1.0, and materials *B* has a relatively higher field strength of 1.2. A large α leads to the non-uniform evaporation rate along the entire surface. As the simulation time step is controlled by the smallest evaporation rate, small time step associated with the large α renders some numerical difficulties. Therefore, α=5 in the exponential evaporation law (Eq. (9)) is chosen to demonstrate the difference between exponential law and linear law where α~1. A large alpha leads to Electrical voltage is applied to the upper and bottom boundary to create a potential field with a vertical gradient of 1.0. A Neumann boundary condition is applied to the left and right side of the computational domain.

Figure 3 compares the dynamic tip shape evolution of the specimen in Figure 2(a) simulated by the level set method and finite difference method [36] for low field material A on top of high field material B. Figures 3(a) and (d) are initial tip shape used in finite difference model and level set method, respectively. Figures 3(b) and (c) are two intermediate tip shapes corresponding to ~32% and ~38% materials evaporated. Material A in the upper layer has a lower evaporation field, or equivalently a higher evaporation rate under the same electric field (higher ratio of applied field to materials evaporation field), leading to a fast evaporation on top. The same trend is also observed in the later stage during evaporation, as shown in Figures 3(c) and (f). Though the current level set model is a planar 2D model in contrast to the true 3D finite difference model by Larson, et al., the preliminary comparison of the tip shape evolution demonstrates a good qualitative agreement between the level set method and the finite difference method.

Figure 4 compares the shape evolution of the tip in Fig. 2(b) (high field material B on top of low field material A) simulated by the level set method and the finite difference method [36]. Since the material on top has a relatively larger field strength, the tip shape is much less flat on top compared to that observed in Figure 3. In contrast, the tip on the axis exhibits a much sharper shape during the evaporation. This is due to the high ratio of the applied field to the material evaporation field strength for the lower layer that leads to the fast evaporation for lower layer material A. The same trend continues during the evaporation process (Figs. 4(c) and (f)). Again, a good qualitative agreement between the level set result and finite difference method is obtained for this configuration.

We have also made a comparison for the third configuration, where the low field material A is on the left side (Fig. 2c). Figure 5 presents the comparison of level set method result with the finite difference result [36]. Figure 5(b) and (d) are the intermediate tip shapes during the evaporation. The initial tip is symmetric, while the evaporated tip is highly asymmetric with respect to the central axis with more material A evaporated than material B due to the different evaporation rates of materials A and B.

**Influence of physical evaporation law on tip shape evolution**

The evaporation law discussed above plays an important role on the tip shape evolution during the evaporation process. Previous calculations have used the exponential evaporation law (Eq. (9)) and get good qualitative agreements with other numerical methods. A plot of the variation of the evaporation rate with the local electric field according to both exponential law (Eq.(9) for α=5) and linear law (Eq.(10)) is presented in Fig. 6. Evaporation rate from linear law is lower than that of the exponential law for relatively smaller electric field ($F_1<F_0$). Otherwise, the exponential law leads to a higher evaporation rate for $F_1>F_0$. Figure 7 compares tip shape evolution calculated by the level set method using two different evaporation laws for three different configurations in Figs. 2 (a), (b) and (c). Figures in the top row are the evaporated shapes using a linear evaporation law while the figures in the bottom row are evaporated shapes using an exponential evaporation law. Due to the fact that the evaporation rate from exponential law is lower than that of the linear law for $F_1<F_0$, materials nearby the vertical edge with smaller local electric field along the normal direction is evaporating much slower using the exponential law. This gives rise to the different tip shapes from two different evaporation laws. Exponential law leads to a sharp and steep vertical edge for all three configurations and resembles the finite difference simulations. Further study on the effect of evaporation laws will be carried out, and experimental results will be utilized for validation in addition to the numerical study.

**CONCLUSIONS**

In our simulations, a level set model has been applied to understand the dynamic tip shape evolution during field evaporation of heterogeneous APT specimens with different geometries. The first example is a composite with spherical particle of higher evaporation field strength embedded in a matrix material of lower evaporation field strength corresponding to the metal nanoparticles supported or embedded in oxides, while the second example represents a bilayer composite made of materials with a 20% difference in evaporation field strength, for which different orientations of the layers within the APT specimen are considered to understand the geometry-dependent specimen shape evolution.

Simulation of evaporated shapes using a level set model shows qualitative agreement with both the finite element and the finite difference method reported in the literature for both examples. In contrast to the "staircase" surface representation in standard finite difference and/or finite element model at the same resolution, the level set model provides a smooth interface representation with inherent sub-grid scale accuracy that leads to the potential improvement in computational efficiency. Effects of two different evaporation laws, i.e. the exponential law and the linear law, on the tip shape evolution were also investigated numerically. The exponential law exhibits better agreement with the finite difference results. Further work will focus on the reconstruction, iterative methods, more rigorous efficiency improvement and evaporation law investigation. Validation with real experimental data is also in progress.

Acknowledgements

The authors acknowledge the funding support from the LDRD-funded Chemical Imaging Initiative at Pacific Northwest National Laboratory, operated for the U.S. Department of Energy by Battelle under contract DE-AC06-76RL01830. A portion of the work was conducted in the William R. Wiley Environmental Molecular Sciences Laboratory (EMSL), a national scientific user facility sponsored by DOE's Office of Biological and Environmental Research located at PNNL, and with the assistance of the William R. Wiley postdoctoral fellowship.

**References:**

[1] A. Devaraj, S. Nag, R. Banerjee, Alpha phase precipitation from phase-separated beta phase in a model Ti-Mo-Al alloy studied by direct coupling of transmission electron microscopy and atom probe tomography, Scripta Mater, 69 (2013) 513-516.
[2] A. Devaraj, S. Nag, B.C. Muddle, R. Banerjee, Competing Martensitic, Bainitic, and Pearlitic Transformations in a Hypoeutectoid Ti-5Cu Alloy, Metall Mater Trans A, 42A (2011) 1139-1143.
[3] A. Devaraj, S. Nag, R. Srinivasan, R.E.A. Williams, S. Banerjee, R. Banerjee, H.L. Fraser, Experimental evidence of concurrent compositional and structural instabilities leading to omega precipitation in titanium-molybdenum alloys, Acta Mater, 60 (2012) 596-609.
[4] A. Devaraj, R.E.A. Williams, S. Nag, R. Srinivasan, H.L. Fraser, R. Banerjee, Three-dimensional morphology and composition of omega precipitates in a binary titanium-molybdenum alloy, Scripta Mater, 61 (2009) 701-704.
[5] S. Nag, A. Devaraj, R. Srinivasan, R.E.A. Williams, N. Gupta, G.B. Viswanathan, J.S. Tiley, S. Banerjee, S.G. Srinivasan, H.L. Fraser, R. Banerjee, Novel Mixed-Mode Phase Transition Involving a Composition-Dependent Displacive Component, Phys Rev Lett, 106 (2011).
[6] H.P. Ng, A. Devaraj, S. Nag, C.J. Bettles, M. Gibson, H.L. Fraser, B.C. Muddle, R. Banerjee, Phase separation and formation of omega phase in the beta matrix of a Ti-V-Cu alloy, Acta Mater, 59 (2011) 2981-2991.
[7] D.N. Seidman, Three-dimensional atom-probe tomography: Advances and applications, Annual Review of Materials Research, 37 (2007) 127-158.
[8] A. Devaraj, R. Colby, W.P. Hess, D.E. Perea, S. Thevuthasan, Role of Photoexcitation and Field Ionization in the Measurement of Accurate Oxide Stoichiometry by Laser-Assisted Atom Probe Tomography, J Phys Chem Lett, 4 (2013) 993-998.
[9] K. Inoue, F. Yano, A. Nishida, H. Takamizawa, T. Tsunomura, Y. Nagai, M. Hasegawa, Dopant distributions in n-MOSFET structure observed by atom probe tomography, Ultramicroscopy, 109 (2009) 1479-1484.
[10] T.F. Kelly, D.J. Larson, K. Thompson, R.L. Alvis, J.H. Bunton, J.D. Olson, B.P. Gorman, Atom probe tomography of electronic materials, Annual Review of Materials Research, 37 (2007) 681-727.
[11] D.E. Perea, E.R. Hemesath, E.J. Schwalbach, J.L. Lensch-Falk, P.W. Voorhees, L.J. Lauhon, Direct measurement of dopant distribution in an individual vapour-liquid-solid nanowire, Nat Nanotechnol, 4 (2009) 315-319.
[12] L.M. Gordon, D. Joester, Nanoscale chemical tomography of buried organic-inorganic interfaces in the chiton tooth, Nature, 469 (2011) 194-197.
[13] L.M. Gordon, L. Tran, D. Joester, Atom Probe Tomography of Apatites and Bone-Type Mineralized Tissues, Acs Nano, 6 (2012) 10667-10675.
[14] J.W. Valley, A.J. Cavosie, T. Ushikubo, D.A. Reinhard, D.F. Lawrence, D.J. Larson, P.H. Clifton, T.F. Kelly, S.A. Wilde, D.E. Moser, M.J. Spicuzza, Hadean age for a post-magma-ocean zircon confirmed by atom-probe tomography, Nat Geosci, 7 (2014) 219-223.
[15] T.F. Kelly, D.J. Larson, Atom probe tomography 2012, Annual Review of Materials Research, 42 (2012) 1-31.
[16] M.K. Miller, K.F. Russell, K. Thompson, R. Alvis, D.J. Larson, Review of atom probe FIB-Based specimen preparation methods, Microsc Microanal, 13 (2007) 428-436.
[17] D.J. Larson, T.J. Prosa, B.M. Geiser, D. Lawrence, C.M. Jones, T.F. Kelly, Atom Probe Tomography for Microelectronics, in: R. Haight, F. Ross, J. Hannon (Eds.) Handbook of Instrumentation and Techniques for Semiconductor Nanostructure Characterization, World Scientific Pub Co Inc, 2012.
[18] K. Thompson, D. Lawrence, D.J. Larson, J.D. Olson, T.F. Kelly, B. Gorman, In situ site-specific specimen preparation for atom probe tomography, Ultramicroscopy, 107 (2007) 131-139.

*[19] P. Bas, A. Bostel, B. Deconihout, D. Blavette, A General Protocol for the Reconstruction of 3d Atom-Probe Data, Appl Surf Sci, 87-88 (1995) 298-304.*
*[20] B.P. Geiser, D.J. Larson, E. Oltman, S. Gerstl, D. Reinhard, T.F. Kelly, T.J. Prosa, Wide-Field-of-View Atom Probe Reconstruction, Microsc Microanal, 15 (2009) 292-293.*
*[21] R. Gomer, Field emission and field ionization, illustrated, reprint ed., American Inst. of Physics1961.*
*[22] B. Gault, M.P. Moody, J.M. Cairney, S.P. Ringer, Atom Probe Microscopy, Springer2012.*
*[23] A. Devaraj, R. Colby, V. Shutthanandan, V. Subramanian, C. Wang, D.E. Perea, S. Thevuthasan, Characterization of embedded metallic nanoparticles in oxides by cross-coupling aberration-corrected STEM and Atom Probe Tomography, Microsc Microanal, 18 (2012) 912-913.*
*[24] A. Devaraj, R. Colby, F. Vurpillot, S. Thevuthasan, Understanding Atom Probe Tomography of Oxide-Supported Metal Nanoparticles by Correlation with Atomic-Resolution Electron Microscopy and Field Evaporation Simulation, J Phys Chem Lett, 5 (2014) 1361-1367.*
*[25] E.A. Marquis, B.P. Geiser, T.J. Prosa, D.J. Larson, Evolution of tip shape during field evaporation of complex multilayer structures, J Microsc-Oxford, 241 (2011) 225-233.*
*[26] J.H. Lee, B.H. Lee, Y.T. Kim, J.J. Kim, S.Y. Lee, K.P. Lee, C.G. Park, Study of vertical Si/SiO2 interface using laser-assisted atom probe tomography and transmission electron microscopy, Micron, 58 (2014) 32-37.*
*[27] D.J. Larson, B. Gault, B.P. Geiser, F. De Geuser, F. Vurpillot, Atom probe tomography spatial reconstruction: Status and directions, Curr Opin Solid St M, 17 (2013) 236-247.*
*[28] F. Vurpillot, B. Gault, B.P. Geiser, D.J. Larson, Reconstructing atom probe data: A review, Ultramicroscopy, 132 (2013) 19-30.*
*[29] D.J. Larson, B.P. Geiser, T.J. Prosa, T.F. Kelly, On the Use of Simulated Field-Evaporated Specimen Apex Shapes in Atom Probe Tomography Data Reconstruction, Microsc Microanal, 18 (2012) 953-963.*
*[30] D.J. Larson, B.P. Geiser, T.J. Prosa, S.S.A. Gerstl, D.A. Reinhard, T.F. Kelly, Improvements in planar feature reconstructions in atom probe tomography, J Microsc-Oxford, 243 (2011) 15-30.*
*[31] D.J. Larson, T.J. Prosa, B.P. Geiser, W.F. Egelhoff, Effect of analysis direction on the measurement of interfacial mixing in thin metal layers with atom probe tomography, Ultramicroscopy, 111 (2011) 506-511.*
*[32] D. Haley, T. Petersen, S. Ringer, G. Smith, Atom probe trajectory mapping using experimental tip shape measurements, J Microsc-Oxford, 244 (2011) 170-180.*
*[33] D. Larson, B. Geiser, T. Prosa, S. Gerstl, D. Reinhard, T. Kelly, Improvements in planar feature reconstructions in atom probe tomography, J Microsc-Oxford, 243 (2011) 15-30.*
*[34] F. De Geuser, B. Gault, A. Bostel, F. Vurpillot, Correlated field evaporation as seen by atom probe tomography, Surface science, 601 (2007) 536-543.*
*[35] B. Gault, M.P. Moody, J.M. Cairney, S.P. Ringer, Analysis Techniques for Atom Probe Tomography, Atom Probe Microscopy, (2012) 213-297.*
*[36] D.J. Larson, B.P. Geiser, T.J. Prosa, T.F. Kelly, On the Use of Simulated Field-Evaporated Specimen Apex Shapes in Atom Probe Tomography Data Reconstruction, Microscopy and Microanalysis, 18 (2012) 953.*
*[37] F. Vurpillot, A. Bostel, D. Blavette, Trajectory overlaps and local magnification in three-dimensional atom probe, Applied Physics Letters, 76 (2000) 3127-3129.*
*[38] C. Oberdorfer, G. Schmitz, On the Field Evaporation Behavior of Dielectric Materials in Three-Dimensional Atom Probe: A Numeric Simulation, Microsc Microanal, 17 (2011) 15.*
*[39] C. Oberdorfer, S.M. Eich, G. Schmitz, A full-scale simulation approach for atom probe tomography, Ultramicroscopy, 128 (2013) 55-67.*
*[40] F. Vurpillot, M. Gruber, G. Da Costa, I. Martin, L. Renaud, A. Bostel, Pragmatic reconstruction methods in atom probe tomography, Ultramicroscopy, 111 (2011) 1286-1294.*

*[41] S.T. Loi, B. Gault, S.P. Ringer, D.J. Larson, B.P. Geiser, Electrostatic simulations of a local electrode atom probe: The dependence of tomographic reconstruction parameters on specimen and microscope geometry, Ultramicroscopy, 132 (2013) 107-113.*
*[42] F. Vurpillot, A. Gaillard, G. Da Costa, B. Deconihout, A model to predict image formation in Atom probe Tomography, Ultramicroscopy, 132 (2013) 152-157.*
*[43] Z. Xu, H. Huang, X. Li, P. Meakin, Phase field and level set methods for modeling solute precipitation and/or dissolution, Computer Physics Communications, 183 (2012) 15-19.*
*[44] Z. Xu, P. Meakin, A phase-field approach to no-slip boundary conditions in dissipative particle dynamics and other particle models for fluid flow in geometrically complex confined systems, J Chem Phys, 130 (2009) 234103.*
*[45] S. Osher, J.A. Sethian, Fronts propagating with curvature-dependent speed: algorithms based on Hamilton-Jacobi formulations, Journal of Computational Physics, 79 (1988) 12-49.*
*[46] J.A. Sethian, Level Set Methods: Evolving Interfaces in Geometry, Fluid Mechanics, Computer Vision, and Materials Science, Cambridge University Press, 1996.*
*[47] Z.J. Xu, K.M. Rosso, S. Bruemmer, Metal oxidation kinetics and the transition from thin to thick films, Phys Chem Chem Phys, 14 (2012) 14534-14539.*
*[48] Z. Xu, K.M. Rosso, S.M. Bruemmer, A generalized mathematical framework for thermal oxidation kinetics, J Chem Phys, 135 (2011) 024108.*
*[49] Z.J. Xu, X. Sun, M.A. Khaleel, A generalized kinetic model for heterogeneous gas-solid reactions, J Chem Phys, 137 (2012).*
*[50] Z. Xu, P. Meakin, Phase-field modeling of solute precipitation and dissolution, The Journal of chemical physics, 129 (2008) 014705.*
*[51] Z. Xu, P. Meakin, Phase-field modeling of two-dimensional solute precipitation/dissolution: Solid fingers and diffusion-limited precipitation, J Chem Phys, 134 (2011) 044137.*
*[52] D. Haley, M.P. Moody, G.D. Smith, Level Set Methods for Modelling Field Evaporation in Atom Probe, Microscopy and microanalysis: the official journal of Microscopy Society of America, Microbeam Analysis Society, Microscopical Society of Canada, (2013) 1-9.*
*[53] J. Sethian, P. Smereka, Level set methods for fluid interfaces, Annual Review of Fluid Mechanics, 35 (2003) 341.*
*[54] L. Arnoldi, E.P. Silaeva, A. Gaillard, F. Vurpillot, I. Blum, L. Rigutti, B. Deconihout, A. Vella, Energy deficit of pulsed-laser field-ionized and field-emitted ions from non-metallic nano-tips, J Appl Phys, 115 (2014).*
*[55] B. Gault, D. Haley, F. De Geuser, M. Moody, E. Marquis, D. Larson, B. Geiser, Advances in the reconstruction of atom probe tomography data, Ultramicroscopy, 111 (2011) 448-457.*
*[56] B.P. Geiser, D.J. Larson, E. Oltman, S.S.A. Gersti, D.A. Reinhard, T.F. Kelly, T.J. Prosa, Wide-field-of view atom probe reconstrcution, Microsc Microanal, 15 (2009) 292-293.*
*[57] F. Vurpillot, A. bostel, D. Blavette, The shape of Æeld emitters and the ion trajectories in three-dimensional atom probes, J Microsc-Oxford, 196 (1999) 332-336.*
*[58] O. Dimond, Modelling the reconstruction of 3D atom probe data [part II thesis], University of Oxford, 1999.*

Figure Captions

Figure 1. Comparison of the tip shape evolution during evaporation simulated from FEM and level set methods. (a) is the initial geometry of the tip with an embedded precipitate. Figures (b) and (d) are simulated tip using FEM when 23% and 25% material evaporated, respectively. Figures (c) and (e) are the corresponding simulated tip using level set method. Figures (f) and (g) show the electrical potential field from (f) FEM and (g) level set method at 23% material evaporated, respectively.

Figure 2. Geometry configurations of the simulation model. Three tips with different configurations are simulated: (a) layered structure material A (low evaporation field strength) on material B (high evaporation field strength), (b) material B on material A, and (c) side-by-side structure (dimension in nm).

Figure 3. Shape evolution of the simulated layer structured tip (Fig 2(a)) with low evaporation field strength material A on top of high evaporation field strength material B. (a)-(c) are simulated by the finite difference method (Larson, et al., 2012a). (d)-(f) are simulated by the level set method. The dash lines in (a)-(c) and the solid lines in (d)-(f) indicate the interface between material A and B.

Figure 4. Shape evolution ion of the simulated layer structured tip (Fig. 2(b)) with high evaporation field strength material B on top of low evaporation field strength material A. (a)-(c) are simulated by the finite difference method (Larson, et al., 2012a). (d)-(f) are simulated by the level set method. The dash lines in (a)-(c) and the solid lines in (d)-(f) indicate the interface between materials A and B.

Figure 5. Shape evolution of the simulated side-by-side structured tip (Fig 2(c)) with low evaporation field strength material A on left of high evaporation field strength material B. (a)-(b) are simulated by the finite difference method (Larson, et al., 2012a). (c)-(d) are simulated by the level set method. The solid lines in (c)-(d) indicate the interface between materials A and B.

Figure 6. Variation of evaporation rates with normalized electrical field for linear (Eq.(9)) and exponential (Eq.(10)) evaporation laws

Figure 7. Tip shape evolution with different geometric configurations using a linear (upper row a-c) and an exponential (Lower row d-f) evaporation law by the level set method. (a)(d) for low filed material A on high field material B in Fig 2(a); (b)(e) for high field material B on low field material A in Fig 2(b); (c)(f) for low field material A on the left side of high field material B in Fig. 2(c). Thick solid lines indicate the position of interface between materials A and B.

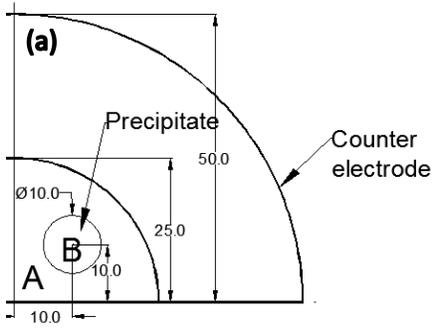

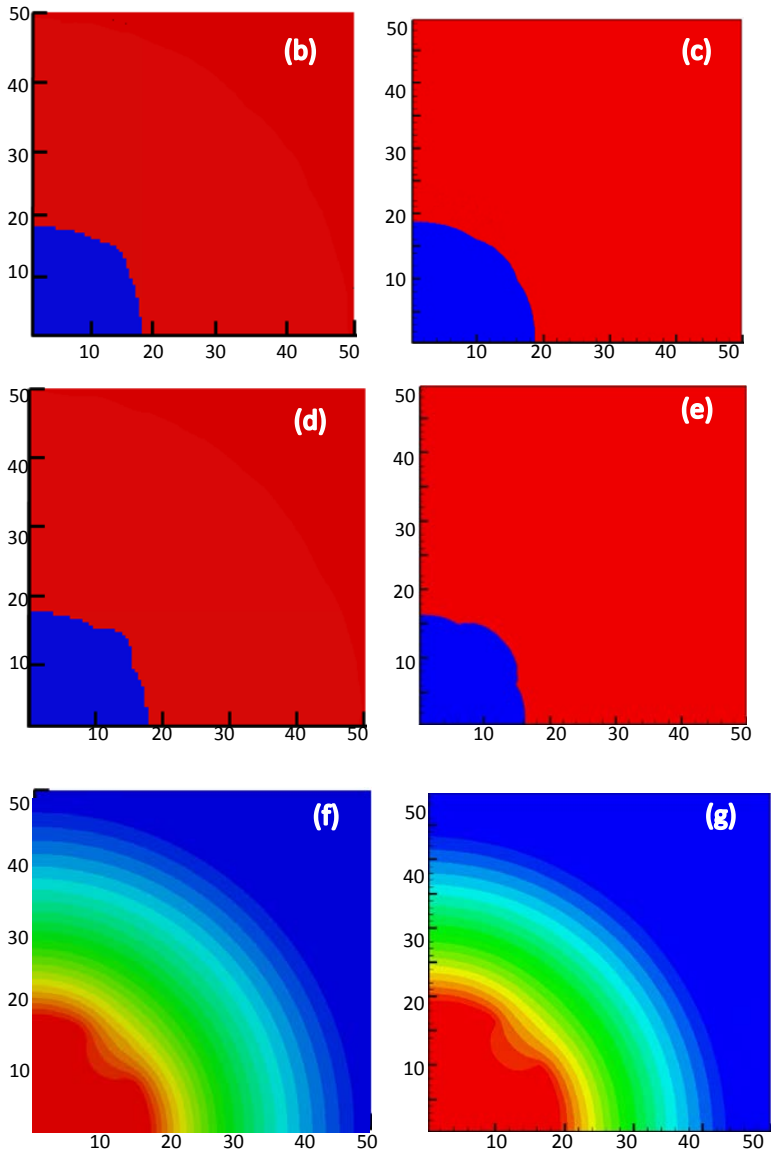

Figure 1. Comparison of the tip shape evolution during evaporation simulated from FEM and level set methods. (a) is the initial geometry of the tip with an embedded precipitate. Figures (b) and (d) are simulated tip using FEM when 23% and 25% material evaporated, respectively. Figures (c) and (e) are the corresponding simulated tip using level set method. Figures (f) and (g)

show the electrical potential field from (f) FEM and (g) level set method at 23% material evaporated, respectively.

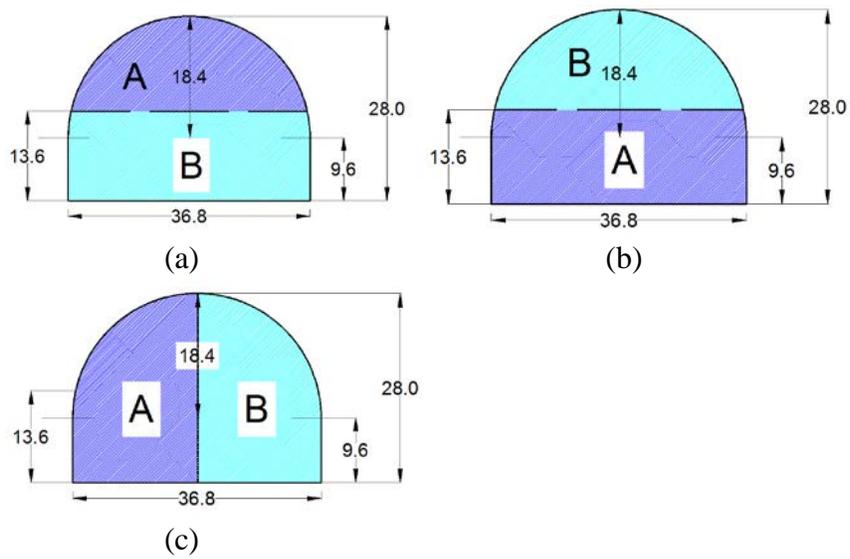

Figure 2. Geometry configurations of the simulation model. Three tips with different configurations are simulated: (a) layered structure material A (low evaporation field strength) on material B (high evaporation field strength), (b) material B on material A, and (c) side-by-side structure (dimension in nm).

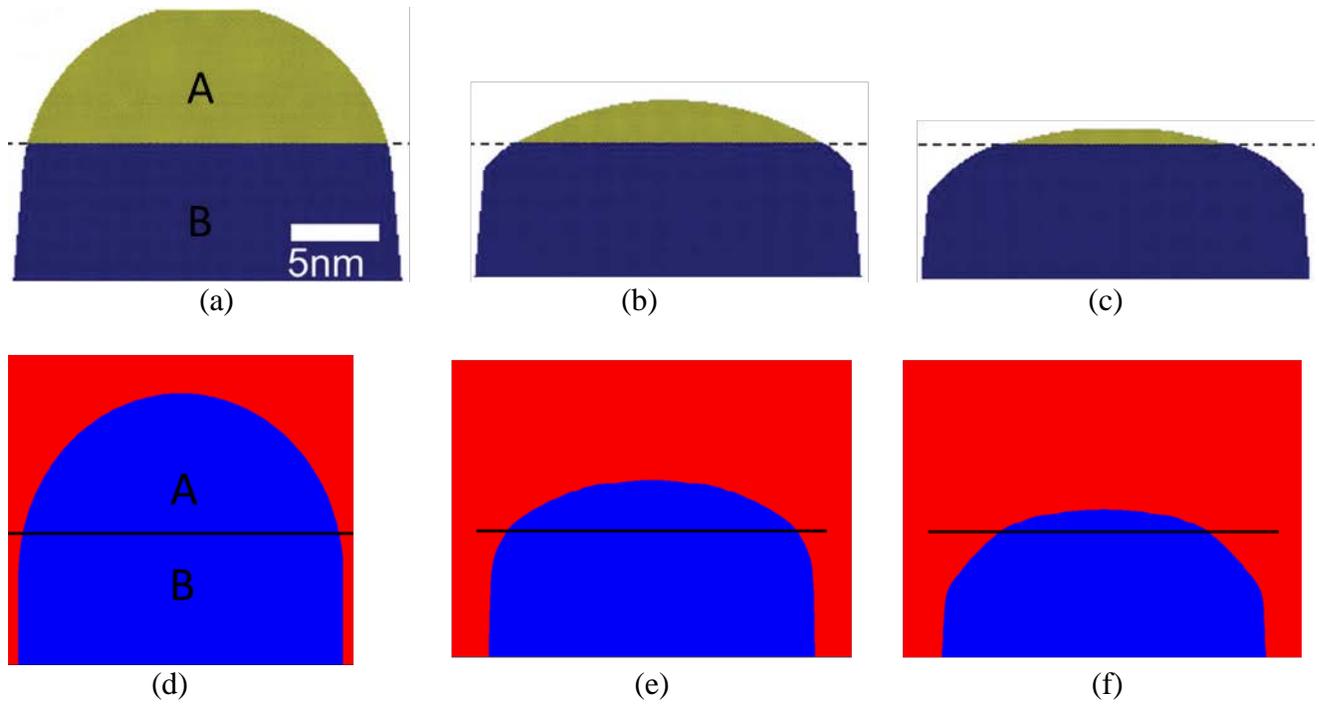

Figure 3. Shape evolution of the simulated layer structured tip (Fig 2(a)) with low evaporation field strength material A on top of high evaporation field strength material B. (a)-(c) are simulated by the finite difference method [36]. (d)-(f) are simulated by the level set method. The dash lines in (a)-(c) and the solid lines in (d)-(f) indicate the interface between material A and B.

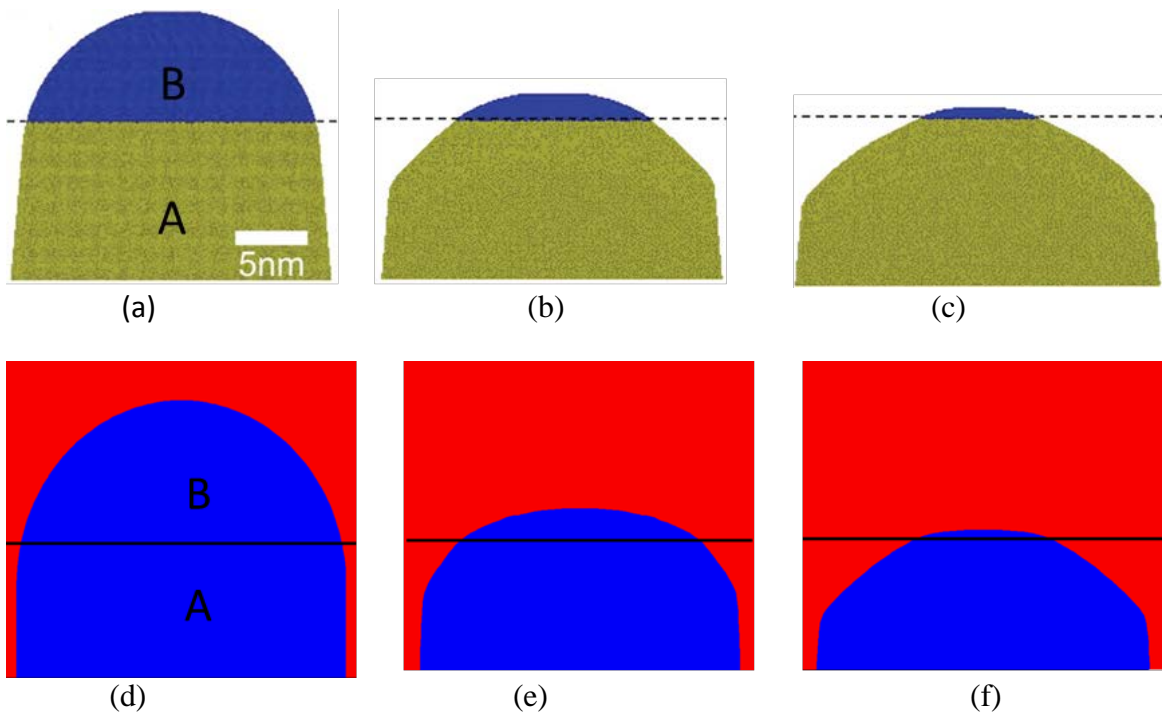

Figure 4. Shape evolution ion of the simulated layer structured tip (Fig. 2(b)) with high evaporation field strength material B on top of low evaporation field strength material A. (a)-(c) are simulated by the finite difference method [36]. (d)-(f) are simulated by the level set method. The dash lines in (a)-(c) and the solid lines in (d)-(f) indicate the interface between materials A and B.

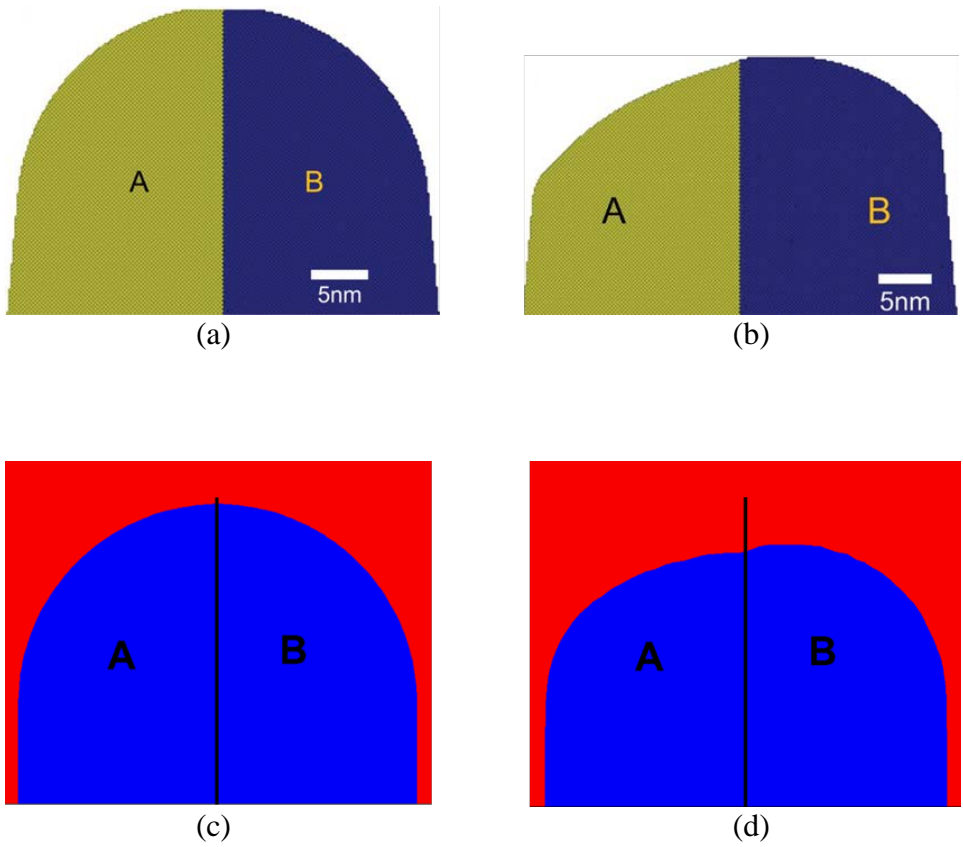

Figure 5. Shape evolution of the simulated side-by-side structured tip (Fig 2(c)) with low evaporation field strength material A on left of high evaporation field strength material B. (a)-(b) are simulated by the finite difference method [36]. (c)-(d) are simulated by the level set method. The solid lines in (c)-(d) indicate the interface between materials A and B.

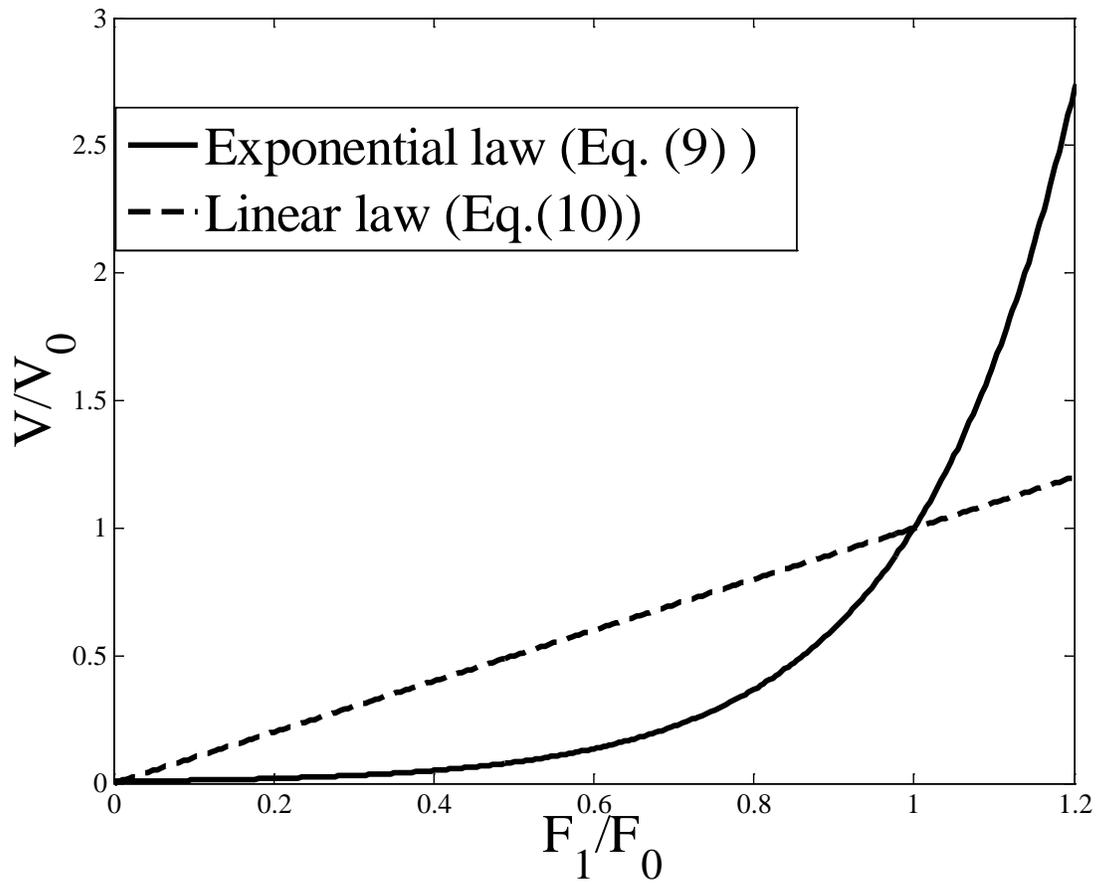

Figure 6. Variation of evaporation rates with normalized electrical field for linear and exponential evaporation laws

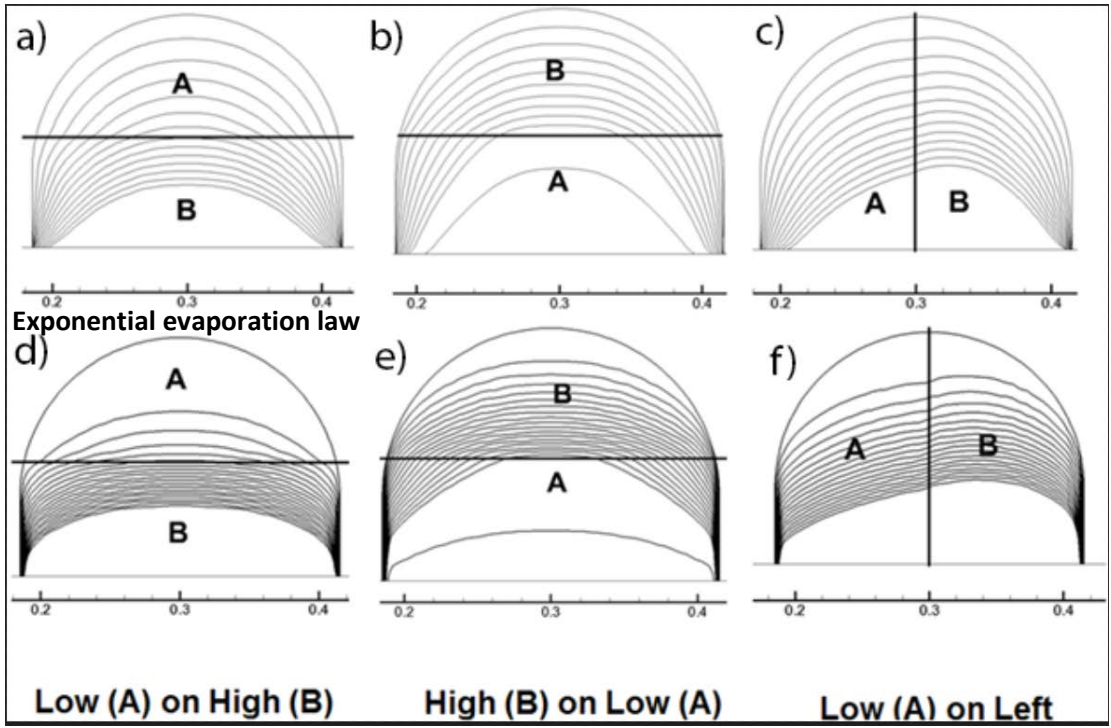

Figure 7. Tip shape evolution with different geometric configurations using a linear (upper row a-c) and an exponential (Lower row d-f) evaporation law by the level set method. (a)(d) for low filed material A on high field material B in Fig 2(a); (b)(e) for high field material B on low field material A in Fig 2(b); (c)(f) for low field material A on the left side of high field material B in Fig. 2(c). Thick solid lines indicate the position of interface between materials A and B.